\documentclass[sn-mathphys]{sn-jnl}% Math and Physical Sciences Reference Style
\jyear{2021}%

\theoremstyle{thmstyleone}%
\theoremstyle{thmstyletwo}%

\theoremstyle{thmstylethree}%

\begin{document}

\title[Assessment of VR interaction methods for neurosurgical data]{Assessment of user-interaction strategies for neurosurgical data navigation and annotation in virtual reality}
%\title[Assessment of neurosurgical VR interaction methods]{Assessment of VR user-interaction strategies for neurosurgical data}

%\title[Article Title] Assessment of user-interaction strategies for neurosurgical data navigation and annotation in virtual reality

\author[1]{\fnm{Owen} \sur{ Hellum}}%\email{iauthor@gmail.com}

\author[1,2]{\fnm{Marta} \sur{ Kersten-Oertel}}%\email{iiauthor@gmail.com}

\author*[1,2]{\fnm{Yiming} \sur{ Xiao}}\email{yiming.xiao@concordia.ca}

\affil[1]{\orgdiv{Department of Computer Science and Software Engineering}, \orgname{Concordia University}, \orgaddress{ \city{Montreal},  \state{Quebec}, \country{Canada}}}

\affil[2]{\orgdiv{PERFORM Centre}, \orgname{Concordia University}, \orgaddress{\city{Montreal}, \state{Quebec}, \country{Canada}}}


%\abstract{The abstract serves both as a general introduction to the topic and as a brief, non-technical summary of the main results and their implications. Authors are advised to check the author instructions for the journal they are submitting to for word limits and if structural elements like subheadings, citations, or equations are permitted.}%

%%================================%%
%% Sample for structured abstract %%
%%================================%%

 \abstract{
 While virtual-reality (VR) has shown great promise in radiological tasks, effective user-interaction strategies that can improve efficiency and ergonomics are still under-explored and systematic evaluations of VR interaction techniques in the context of complex anatomical models are rare. Therefore, our study aims to identify the most effective interaction techniques for two common neurosurgical planning tasks in VR (point annotation and note-taking) from the state-of-the-arts, and propose a novel technique for efficient sub-volume selection necessary in neuroanatomical navigation. We assessed seven user-interaction methods with multiple input modalities (gaze, head motion, controller, and voice) for point placement and note-taking in the context of annotating brain aneurysms for cerebrovascular surgery. Furthermore, we proposed and evaluated a novel technique, called \emph{magnified selection diorama (Maserama)} for easy navigation and selection of complex 3D anatomies in VR. Both quantitative and semi-quantitative (i.e., NASA Task Load Index) metrics were employed through user studies to reveal the performance of each interaction scheme in terms of accuracy, efficiency, and usability. Our evaluations demonstrated that controller-based interaction is preferred over eye-tracking-based methods for point placement while voice recording and virtual keyboard typing are better than freehand writing for note-taking. Furthermore, our new \emph{Maserama} sub-volume selection technique was proven to be highly efficient and easy-to-use. Our study is the first to provide a systematic assessment of existing and new VR interaction schemes for neurosurgical data navigation and annotation. It offers valuable insights and tools to guide the design of future VR systems for radiological and surgical applications.

}




\maketitle

\section{Introduction}\label{sec1}
Medical imaging techniques, such as magnetic resonance imaging (MRI) and computed tomography (CT) have become integral components of clinical practices. They allow the clinicians to inspect the bodily tissues and pathologies (e.g., tumors) non-invasively to facilitate diagnosis and planning of surgical treatment. With the advancement of bioimaging techniques, newer specialized imaging contrasts, such as diffusion MRI are being introduced to improve the visualization of specific anatomies, pathologies, and physiological processes, and thus an increasing amount of imaging data are required to provide the comprehensive view needed for high-quality clinical diagnoses and surgical planning. This is especially true for neurosurgical care. However, traditional 2D display-based radiological and surgical planning software poses challenges for the clinicians to understand, navigate, and annotate complex 3D anatomical data ~\cite{bib1,bib33}, as well as to conduct multi-site and multi-disciplinary collaboration, which is necessary for difficult clinical cases and to improve accessibility of healthcare. This is due to a growing dependence on the valuable data provided by modern imaging techniques, which can lead to an imperfect or overly-complicated representation when shown in 2D. Virtual reality (VR) can address these challenges through more intuitive visualization and interaction of 3D data in an immersive environment with the use of a head-mounted display (HMD). 

So far, medical VR has demonstrated great potential in clinical practices and medical education. More specifically for neurosurgical applications ~\cite{bib2}, there have been a number of systems that were introduced for different neurosurgical procedures and purposes, with innovations in methodologies and applications. A recent comprehensive review on virtual reality for neurosurgical applications has been conducted by Fiani et al ~\cite{bib2}. With the capacity to provide an immersive environment, neurosurgical VR has been adopted in surgical rehearsal and training to minimize risks during the interventions. For example, Alaraj et al. ~\cite{bib3} employed a virtual reality system with real-time haptic feedback to simulate the cerebral aneurysm clipping procedure. Heredia-Pérez et al ~\cite{bib4} built a VR system to train physicians for robotic brain tumor surgery. In comparison to neurosurgical training and related education, only a few neurosurgical planning systems have been reported. With video game controllers, Shao et al. ~\cite{bib5} employed VR visualization to improve the quality of surgical plans for skull base tumor resection. Kockro et al ~\cite{bib6} reported a large-scale study with excellent clinical outcomes for a VR-based surgical planning system to allow clip pre-selection and positioning. Most recently, Hellum et al. ~\cite{bib7} proposed the first VR system to plan the deep brain stimulation procedure for treating Parkinson’s disease. With a voodoo doll interaction technique to locate the stimulation target and gaze-tracking for surgical trajectory planning, the system achieves a surgical plan within ~1 min. Finally, VR has also been used for patient engagement in neurosurgical consultation to relieve their stress and facilitate decision making through intuitive visualization of the pathologies and surgical plans. In 2018, Collins et al. ~\cite{bib8} conducted one such pilot study for deep brain stimulation surgery with positive feedback and later Louis et al. ~\cite{bib9} examined the impact of VR-based neurosurgical consultation, demonstrating improved patient satisfaction. Despite the current progress in medical VR, most literature on medical VR~\cite{bib2} only focuses on qualitative spatial understanding of the pathology and anatomy (e.g., brain tumors), instead of developing and evaluating human computer interaction (HCI) schemes that can improve the efficiency, accuracy, and usability of the relevant VR systems for applications (e.g., surgical planning) that demand high accuracy.

To date, a number of HCI methods in VR have been proposed~\cite{bib10} and demonstrated mainly through interacting (e.g., pointing and selection) with simplistic and artificial objects/scenes (e.g., cubes or spheres). Besides the staple hand-held controllers, eye-tracking technology has recently been integrated into commercial VR HMDs. This can potentially offer richer HCI methodologies to improve efficiency and ergonomics, but gaze-based HCI strategies have been rarely assessed in 3D interaction~\cite{bib11, bib12}, where depth information and bigger interaction space can pose additional challenges compared to  simpler 2D menus~\cite{bib10}. This is especially true for medical VR in radiological applications, where the target virtual objects (e.g., blood vessels and pathologies) often have complex geometries, small sizes, and elaborate spatial compositions in contrast to the simplistic artificial objects commonly used in VR testing. To this end, comprehensive evaluation and comparison of VR user-interaction methods, particularly those based on eye-tracking technology have not been conducted with the consideration of these unique properties of anatomical models. Furthermore, with the need to improve the accessibility and knowledge sharing of clinical expertise and medical training, especially in the post-pandemic era, there is a growing interest in collaborative medical VR, which is shown to improve the quality of diagnosis ~\cite{bib13}, surgical planning ~\cite{bib14}, and clinical education ~\cite{bib15}. Effective user-interaction techniques, especially anatomical annotation, note-taking, and object selection are expected to ensure the efficiency, clarity, and accuracy of the communication in a multi-user VR environment, whether synchronous or asynchronous. Specifically, some earlier explorations have demonstrated the clinical benefits of appropriate task-specific object annotation ~\cite{bib16} and commentary ~\cite{bib15} for improved surgical planning workflow efficiency and understanding of the anatomy.

The objectives of the research are to identify the most effective state-of-the-art interaction techniques for two common neurosurgical navigation and annotation tasks in VR, including 3D point annotation and note-taking, and to propose a new sub-volume selection technique for navigating complex neuroanatomical data. More specifically, we performed three sets of user studies for the tasks of localizing (3D point placement) and annotating (note-taking) cerebrovascular abnormality (i.e., aneurysms) and selecting neuroanatomical structures for surgical planning (sub-volume selection). First, for point placement, which is a basic and key task to allow incision point annotation ~\cite{bib16}, distance measurement and trajectory planning ~\cite{bib7}, we compared eye-tracking-based (i.e., \emph{Eye-dwell} and \emph{Gaze \& Controller}), controller-based, and head-pointing-based methods. Second, for commentary annotation, we evaluated three techniques: \emph{Voice recording}, \emph{Handwriting}, and \emph{Virtual keyboard typing}. Similar techniques have been implemented before in collaborative interaction for sharing information and calling attention to certain points of interest in surgical applications ~\cite{bib14,bib16}. Finally, we proposed and validated a novel volume-of-interest (VOI) selection strategy for exploring and navigating 3D anatomical models with complex composition, called magnified selection diorama or \emph{Maserama}. With the assumption that eye-tracking-based techniques and voice recording can offer more intuitive and ergonomic interaction than their other counterparts by design, we expect that they are more optimal interaction strategies in 3D point placement and note-taking, respectively. In addition, we hypothesize that the newly proposed \emph{Maserama} sub-volume selection technique is efficient and easy-to-use. The knowledge drawn from the study provides key insights for the recommendation of user-interaction methods for medical VR, and together with the newly proposed \emph{Maserama} technique, which is the first technique of its kind for medical VR HCI, can form an efficient workflow for precision-driven neuroanatomical navigation and annotation.

\section{Methods and Materials}\label{sec2}
We employed two groups of operations to evaluate and compare different VR interaction techniques for navigating and annotating 3D brain models. In the first group, point placement and the associated commentary annotation were performed for brain aneurysms, which are balloonings of weakened cerebral arterial walls and are usually treated with surgical clipping or endovascular coiling~\cite{bib3}. In the second group, we assessed the newly proposed anatomical navigation strategy, \emph{Maserama} in selecting small subcortical structures that are situated deep in the brain model. The full workflows and setups of the experiments are detailed in Fig.S1-S4 of the \emph{Supplementary materials}.

\subsection{Neurological model building}
We constructed the virtual brain model (see Fig.1) for our study with structural segmentations of 3T MRI scans of a healthy subject from the TubeTK database~\footnote{https://public.kitware.com/Wiki/TubeTK/Data}, which contains matching T1w and T2w structural MRI (1x1x1 $mm^{3}$ resolution) and MR angiography (MRA, 0.5x0.5x0.8 $mm^{3}$ resolution) for each subject. The base virtual model was made of four sets of anatomical structure, including the cerebral vasculatures, lateral ventricles, subcortical nuclei, and brain surface. To segment them, we first extract a brain mask with the BEaST algorithm~\cite{bib17} from the T1w MRI. The brain vasculatures were segmented using the Frangi vesselness filter~\cite{bib18} and then further refined manually by the author YX using ITK-SNAP (http://itksnap.org). Eleven subcortical nuclei were segmented by warping high-resolution labels from the MNI-PD25 atlas \cite{bib19} to the subject’s anatomy. Lateral ventricles were extracted with the SNAKE tool in ITK-SNAP. Lastly, all segmentations were saved as polygon mesh models in .obj formats to build the 3D scene. To test VR interaction methods for 3D point placement and the associated note-taking, we added “artificial aneurysms” as small spheres of diameter=3mm at different locations of the blood vessel model (see Fig 1).

\begin{figure}[h]%
\vspace{-0.5cm}
\centering
\includegraphics[width=1\textwidth]{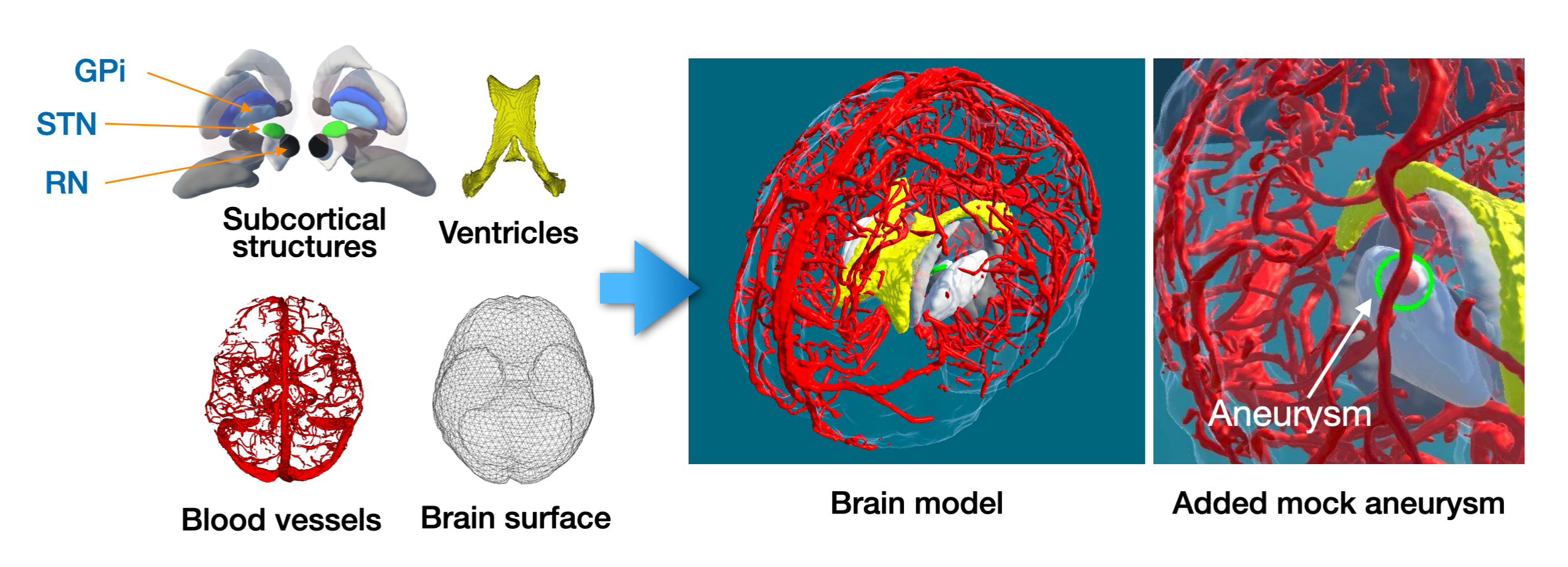}
\caption{Demonstration of the brain model and artificially augmented aneurysms.}\label{fig1}
\vspace{-0.3cm}
\end{figure}

\subsection{Virtual reality environment construction}
To build the VR system, we employed the HTC VIVE Pro Eye VR HMD that integrates eye-tracking capabilities and its associated hand-held controllers. The virtual environment was created using the Unity game engine software (Version 2020.3.8f1) and SteamVR on a Razer Blade 15 laptop (Intel Core i7 CPU, NVIDIA GeForce RTX 2070 GPU, and 16.0 GB RAM). Different neuroanatomical structures were assembled to form the virtual brain model with distinct color coding and the brain surface model was rendered as a semi-transparent layer to allow visualization of the internal structures while offering a sense of the brain volume. We positioned the brain model in the virtual space in a “museum exhibition mode” for active exploration, thus simplifying the operation of the system. No lagging or frame freezing was observed for our system.

\subsection{User interaction strategies and experimental setup}
\subsubsection{Point placement methodologies}
Accurate and efficient 3D point placement is instrumental in various tasks in medical VR that demand high precision, including size measurement and localization of pathology or incision points~\cite{bib16}. To explore the optimal method of point placement, we tested 4 distinct strategies, including \emph{Controller-only}, \emph{Head pointing and Controller}, \emph{Gaze and Controller}, and \emph{Eye-dwell}, with the last two utilizing eye-tracking. These strategies are demonstrated in Fig.2. In \emph{Controller-only} point placement,  a small sphere is fixed to the end of a virtual rod, which is rigidly attached to the end of the right controller. Pressing the right controller’s trigger button creates an instance of the sphere at the location that it points to. As careful visual inspection often requires drawing the head closer to the region of interest, in \emph{Head pointing \& Controller}, a point is statically affixed in front of the participant’s head. The participant moves their head to position the point where needed and uses the right trigger to confirm the placement. The third strategy, \emph{Gaze \& Controller}, uses tracked gaze location on the 3D model to determine the point position, which is confirmed by pressing the right controller’s trigger. In the final strategy, \emph{Eye-dwell}, instead of using the controller, the point placement location is confirmed by holding the gaze at the same location beyond a predetermined time interval, which was set at 700ms for our experiment to balance accuracy (i.e., mistaking short gaze rest as point placement) and efficiency. To further improve the user experience, we also designed a novel feedback mechanism for signaling the remaining time for eye-dwells. Specifically, a halo-like outline was shown around the participant’s gaze point with its width corresponding to a timer. The timer decreases when the participant fixed the  gaze, but resets when the participant looks away from that location. When the timer (and the corresponding halo outline width) reaches zero, a point would be placed at the designated location. This strategy doesn't require controllers and has not been extensively tested for 3D objects in VR.

For each strategy, the point placement could be revised at will. After the participant was satisfied with the location of each placement, they confirmed it by pressing the left controller’s trigger. In our experimental setup, all the strategies were presented in a random order for each participant, and for each strategy, point placements were performed to localize four aneurysms, shown as red and transparent spheres at various locations of the blood vessels in the frontal lobe bilaterally. All aneurysm locations are shown in Fig. S5 of the \emph{Supplementary materials} for both the Point Placement (4 locations) and Annotation tasks (2 locations). To guide the participants, a cyan outline around the aneurysm was used to indicate the current one to tag.

\begin{figure}[h]%
\vspace{-0.5cm}
\centering
\includegraphics[width=\textwidth]{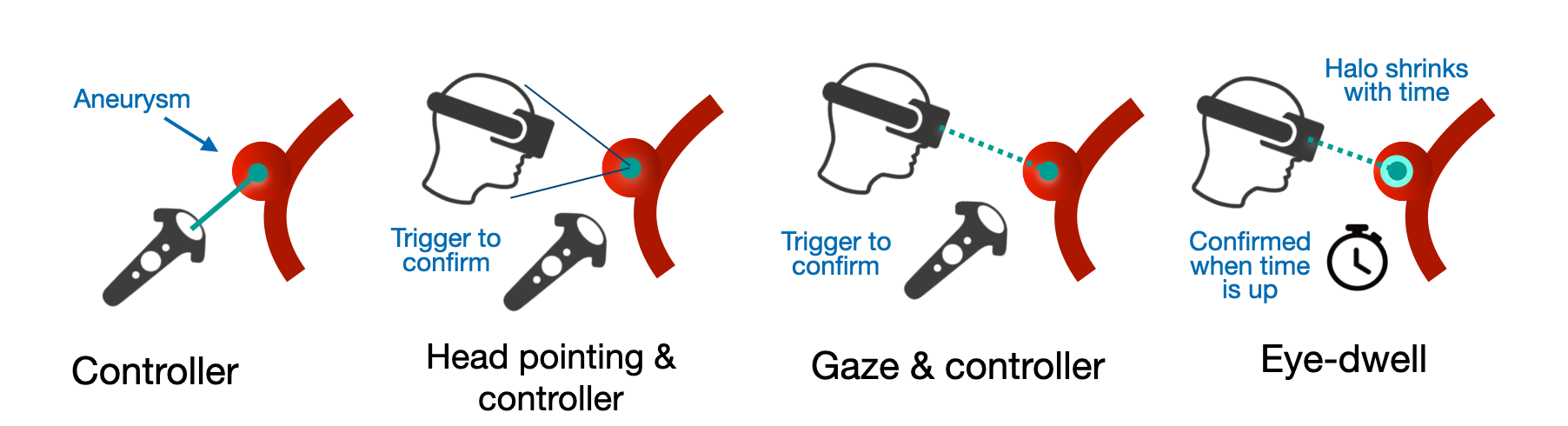}
\caption{Demonstration of four different VR interaction strategies for point placement. From left to right: controller-based, head pointing \& controller, gaze \& controller, and eye-dwell.}\label{fig2}
\vspace{-0.3cm}
\end{figure}

\subsubsection{Annotation methodologies}
Adding notes or commentary regarding specific locations of an anatomical model can greatly facilitate asynchronous clinical collaboration within VR~\cite{bib14}. Following point placement using the controller-based method to tag an aneurysm, a note was created and attached to this location. For this task, three note-taking strategies were tested (see Fig.3), including \emph{Handwriting}, \emph{Virtual keyboard} and \emph{Voice recording}. Specifically, \emph{Handwriting} displays a large 2D board which is rigidly attached to the left controller. While holding on the trigger, the participant uses the right controller as a virtual pen to write on this board’s surface. Any text written this way may be erased and rewritten. On the other hand, \emph{Virtual keyboard} uses a 2D keyboard attached to the left controller, and the participant uses the right controller to “type” on it by pointing with a virtual rod and confirming the key with the right trigger. Finally, \emph{Voice recording} presents the user with an audio recording interface attached to the left controller with buttons for recording and playing back messages. The participants’ voices were recorded through the VR headset.

During the experiment, for each strategy, the target point could be repositioned using the controller, and each note could be redone till satisfaction. After the participant was satisfied with the point placement and the note, they confirmed each separately by pressing the left controller’s trigger. Again, each strategy was presented in a random order for the participants to avoid potential preferences due to their orders of appearance, and for each strategy, the task was performed for two different locations of aneurysms. To examine different demands for note taking, we designed a number of short and long messages, where the short ones are 2-to-3-words long while the long messages entailed full sentences. The full collection of phrases are detailed in Table S1 of the \emph{Supplementary Materials}. Each participant was required to annotate one short and one long note randomly selected from the collection of phrases. To guide the participants, the text to be annotated would appear above the brain model, and they would write, type, or record each respective note.

\begin{figure}[h]%

\centering
\includegraphics[width=\textwidth]{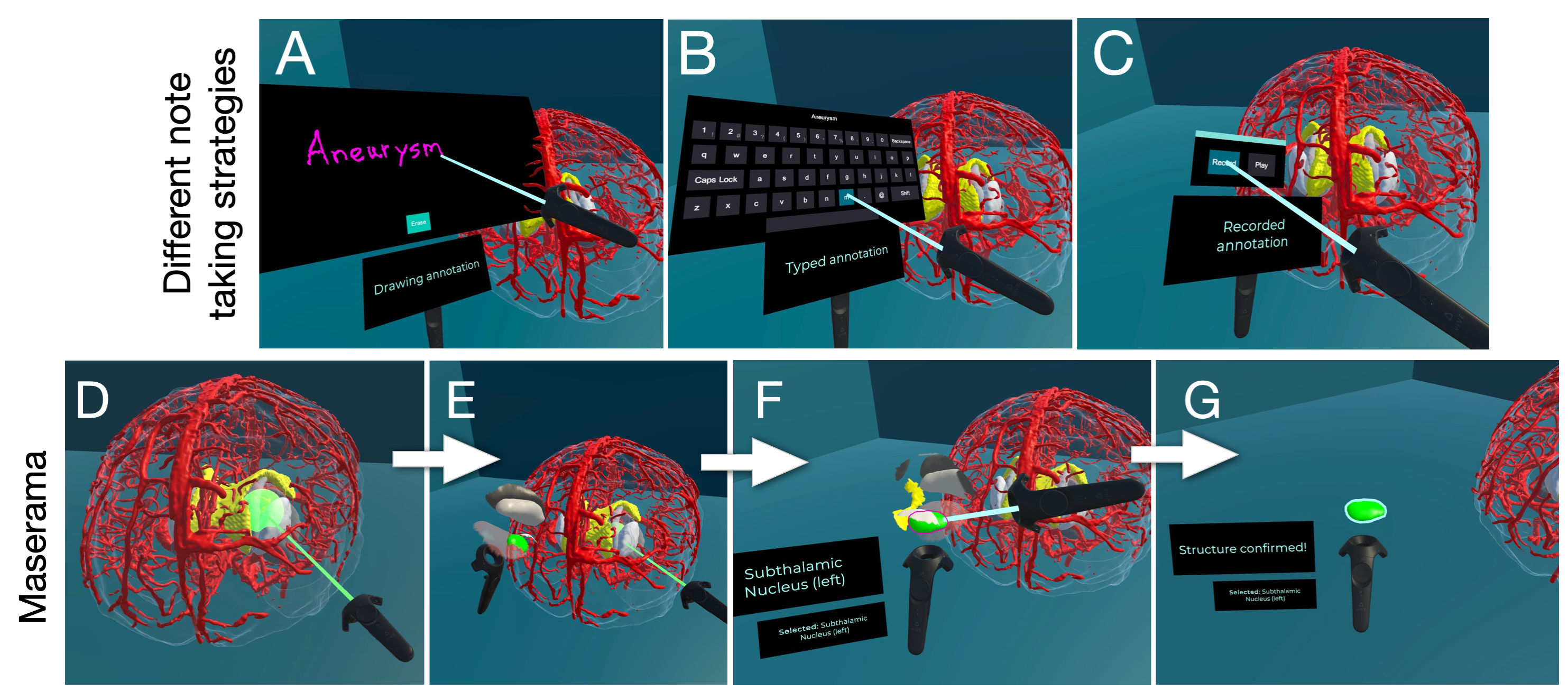}
\caption{Demonstration of different note-taking strategies (top) and workflow of Maserama selection of the left STN (bottom). A = \emph{Handwriting}; B = \emph{Virtual keyboard}; C = \emph{Voice recording}. From D to G: sub-volume exploration, confirmation of sub-volume, exploration of anatomical structures with annotations, and confirmation of target structure selection.}\label{fig3}

\end{figure}

\begin{figure}[h]%
\centering
\includegraphics[width=\textwidth]{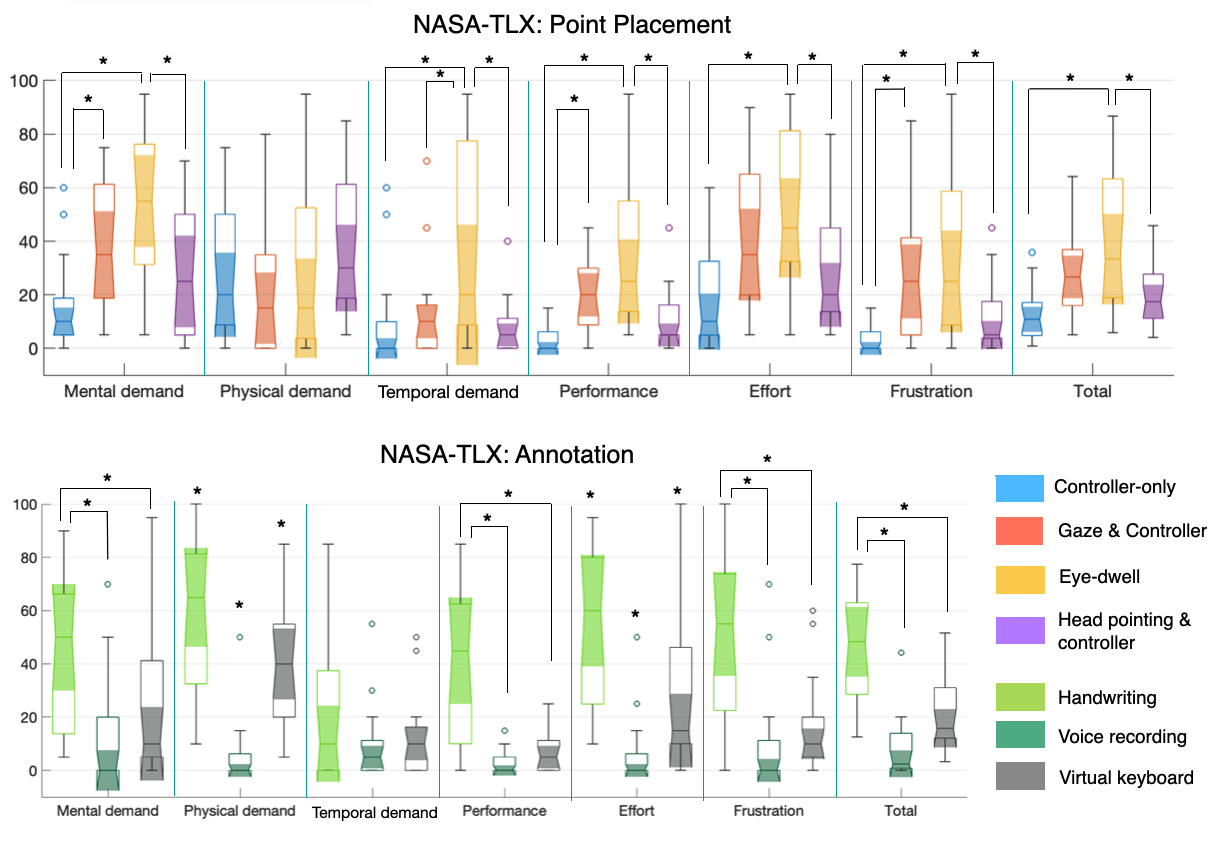}
\caption{Boxplots of NASA Task Load Index (NASA-TLX) measures for both Point Placement and Annotation tasks. Pairwise differences with statistical significance (p$<$0.05) are marked with connected lines and an asterisk sign on top. When all items are significantly different from each other, only the asterisk signs are indicated on all associated boxplots.A low score in each item means a more favorable evaluation}\label{fig5}
\end{figure}

\subsubsection{Maserama selection}
Anatomical models such as brain structures are often layered upon each other with complex configurations (Fig.~\ref{fig1}). In real-time collaboration, it is beneficial to allow individual parties to navigate sub-volumes of the model for closer inspection while keeping the main model intact. To achieve this, we designed a light-weight solution called \emph{Maserama}, which uses a large sphere with an extension rod attached to the right controller to “scoop” a sub-volume for detailed navigation and structure selection. This is achieved in 3 steps (Fig.3E-3G). First, the participant used the right controller to position the selection sphere within the composition of structures, and then pressed the right trigger to confirm the selection. This way, a magnified replica of the selected structures intersecting with the spherical volume will be attached to the left controller for closer inspection. Second, after confirming the cropping with the left trigger, the participant could then select one of the structures in the sub-volume using the right controller as a laser pointer. Two small text boxes attached to the left controller displayed the current structure that the right controller was pointing towards, and the structure that was selected. This offered further information to help navigate the anatomy. Lastly, when the participant was satisfied with their selection, they confirmed it by pressing the left trigger. Only the selected structure is left for further manipulations.

For the experiment, as demonstrated in Fig.3, we requested the participant to slice out a sub-volume at the center of the brain model and select one of the subcortical structures: the subthalamic nucleus (STN), globus pallidus interna (GPi), and red nucleus (RN). Both the STN and GPi are popular surgical targets in deep brain stimulation therapy to treat Parkinson’s disease~\cite{bib20}.

\subsubsection{User study design and system validation}
The study was approved by the Research Ethics Board of Concordia University (Protocol NO.: 30015332) and all subjects gave their consent before their participation. We recruited 17 subjects (age=30.2±5.3 yo, 4 female, 13 male) to participate in the user study. The participants were invited via departmental emails and were not compensated monetarily for their involvement. Subjects with glasses were allowed to keep them on when participating as the VR headset is compatible with them. No one experienced VR sickness. The participants were first given a hands-on tutorial with the VR system for each user interaction strategy. This helped familiarize themselves with the virtual environment, calibrate the eye-tracking system, and practice the respective tasks. In the tutorial, participants were presented with all strategies for point placement and note-taking, and can tag and annotate any points in the brain model. In this phase, no aneurysms or pre-defined phrases were shown to them. For the \emph{Maserama} selection technique, participants were allowed to test out the spherical slicing to freely select any regions and structures of the virtual brain. All participants completed the tutorial within 5 min for all interaction strategies. In the “study” phase, the relevant modules were loaded in the order of \emph{Point Placement}, \emph{Note-taking}, and \emph{Maserama Selection}. After finishing each module, the participants would complete their assessments of different strategies, and then proceed to the next module till full completion.

To validate our proposed VR system, we employed both quantitative and semi-quantitative measures. In regards to the former, relevant quantitative metrics were gathered directly from within the environment during each of the three tasks. Specifically, data were gathered regarding individual completion times (in seconds), number of replacements before confirmation, and accuracy to targets (in mm) where applicable. For the \emph{Point Placement} and \emph{Annotation tasks}, these metrics were gathered for each of their respective types. Relevant results were also compared between long and short phrases for annotation.

To verify the usability of each interaction method, we employed the NASA Task Load Index (TLX)~\cite{bib21}. The evaluation consists of questions pertaining to mental, physical, and temporal demand, as well as performance (i.e., how well a task is completed), effort and frustration factors. A low score in each item means a more favorable evaluation. These questions are rated on scales of 1 to 21, and the final score for each sub-category is computed as “(score -1)*5” so that it is normalized to the range of [0,100] for easy understanding. Finally, the total NASA TLX score is computed as the average of all 6 sub-scores (range=[0, 100]), allowing us to compare the level of task load across different tasks. Within the tasks of \emph{Point Placement} and \emph{Annotation}, we performed a one-way analysis of variance (ANOVA) and post-hoc multiple comparison (Tukey) tests to compare different strategies in the measures of completion times and NASA TLX. Finally, we asked the participants to provide their familiarity with VR technology and brain anatomy, as well as a few general questions on their preferred strategy in each task and how to improve the system. Among the cohort, 14 were familiar or somewhat familiar with VR and general brain anatomy. All participants were right-handed and not colorblind.

\section{Results}\label{sec3}
\subsection{Quantitative evaluation of user interaction strategies}
The accuracy and time metrics from the participants were collected from the VR application for analysis. Note that data related to accuracy were only collected during point placement tasks. Here, the point placement accuracy is defined as the Euclidean distance between the tagged point and the closest point within and on the surface of the “artificial” aneurysm. Point Placement techniques that had the ability to place points beneath the aneurysm’s surface, namely \emph{Controller-only} and \emph{Head pointing \& Controller}, were evaluated with a mean accuracy of 0 mm. Mean accuracy for point placement was consistently on or within the “artificial” aneurysm’s surface for \emph{Controller-only} and \emph{Head pointing \& Controller}, and was within about 0.1 mm for \emph{Gaze \& Controller} and \emph{Eye-dwell}. For task completion time, 3 of 4 strategies for point placement took under 15 sec, and the annotation methods took 25-42 sec. For each task, the completion times didn’t differ significantly among the relevant methods (p$>$0.05). Lastly, the completion time for \emph{Maserama} was 21.0±9.3 sec. The full quantitative measures are detailed in Table 1.

\subsection{Task load evaluation}
The obtained NASA TLX of each category and the total score are illustrated as boxplots in Fig.4 for the tasks of \emph{Point Placement} and \emph{Annotation}. For \emph{Point Placement}, \emph{Controller-only} performs significantly better (p$<$0.05) than the two eye-tracking-based techniques in the categories of mental demand, performance, and frustration while \emph{Controller-only} and \emph{Head pointing \& Controller} are comparable (p$>$0.05) in these categories. In terms of physical demand, there is no significant difference between the four methods, and \emph{Eye-dwell} requires the highest temporal demand and more effort than non-eye-tracking methods for Point Placement (p$<$0.05). Finally, when assessing the total NASA TLX, \emph{Eye-dwell} is inferior to both \emph{Controller-only} and \emph{Head pointing \& Controller} (p$<$0.05) with \emph{Controller-only} having the best mean total score. Interestingly, \emph{Gaze \& controller} is comparable to the rest of the techniques (p$>$0.05). 

When it comes to note-taking, the results for physical demand and effort are distinct among all three techniques (p$<$0.05), with the ranking from the best to the worst being \emph{Voice Recording}, \emph{Virtual keyboard}, and \emph{Handwriting}, but their temporal demand do not show significant differences (p$>$0.05). In the categories of mental demand, performance, and frustration, \emph{Handwriting} offers inferior results than the rest, which demonstrate comparable results in these categories. Lastly, for the total NASA TLX, \emph{Voice recording} and \emph{Virtual keyboard} are better than \emph{Handwriting}, but their between-group difference is not significant (p$>$0.05), with the former having a lower average total score.

\emph{Maserama} was evaluated individually for the task of sub-volume selection. The scores for the sub-categories of NASA TLX are 15±14.7 (mental demand), 14.4±16.8 (physical demand), 5.6±8.6 (temporal demand), 5.6±8.6 (performance), 11.5±12.5 (effort), 9.1±15.5 (frustration). The total score is computed as 10.2±9.9. The NASA TLX scores are illustrated in Fig. 5. The values are in general on par with \emph{Controller-only} for \emph{Point Placement} task and \emph{Voice recording} for \emph{Annotation}. 

\begin{figure}
\centering
\includegraphics[width=\textwidth]{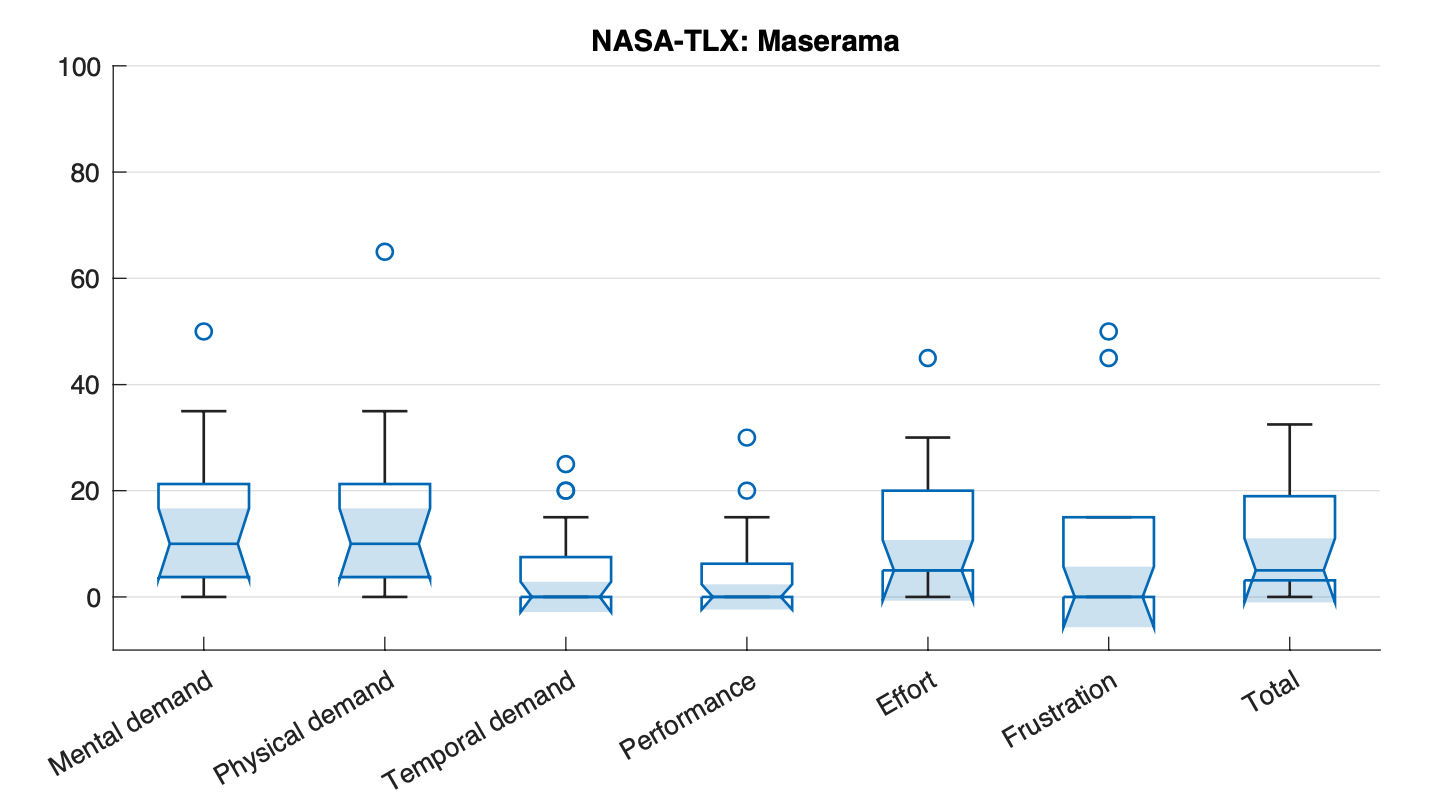}
\caption{Boxplots for NASA TLX measures for Maserama sub-volume selection.}

\end{figure}

%\usepackage{booktabs}
\begin{table}[]
\resizebox{\textwidth}{!}{\begin{tabular}{@{}c|ccccc@{}}
\toprule
\textbf{General test} & \textbf{Sub-task} & \textbf{Completion time(s)} & \textbf{Accuracy (mm)} & \textbf{Replacement} \\ \midrule
Point placement & Controller-only & 14.3±11.7 & \begin{tabular}[c]{@{}c@{}}0.0±0.0\\ (0.4±0.2 to COM)\end{tabular} & 0.8±1.1 \\
\\
 & Head pointing \& Controller & 14.0±9.5 & \begin{tabular}[c]{@{}c@{}}0.006±0.02\\ (0.5±0.2 to COM)\end{tabular} & 0.3±0.4 \\
 \\
 & Gaze \& Controller & 20.2±15.5 & \begin{tabular}[c]{@{}c@{}}0.1±0.1\\ (1.6±0.1 to COM)\end{tabular} & 2.1±1.9 \\
 \\
 & Eye-Dwell & 14.1±7.7 & \begin{tabular}[c]{@{}c@{}}0.1±0.2\\ (1.5±0.3 to COM)\end{tabular} & 5.8±4.5 \\
 \\
 \hline
 \\
 
Annotation & Handwriting (Long Phrases) & 40.0±19.8 & N/A & 0.1±0.2 \\
\\
 & Handwriting (Short Phrases) & 27.4±9.2 & N/A & 0.1±0.2 \\
 \\
 & Typed (Long Phrases) & 41.7±17.3 & N/A & 0.2±0.2 \\

\\
 & Typed (Short Phrases) & 25.8±6.2 & N/A & 0.2±0.3 \\
\\

 & Voice Recording (Long Phrases) & 32.4±12.1 & N/A & 0.1±0.2 \\
\\
 & Voice Recording (Short Phrases) & 32.0±14.1 & N/A & 0.2±0.2 & \\

 \hline
Sub-volume selection & Maserama & 21.0±9.3 & N/A & \begin{tabular}[c]{@{}c@{}}Cluster selection: 0.4±0.3\\ Structure selection: 0.0±0.1\end{tabular} \\ \bottomrule

\end{tabular}}

\caption{Quantitative data gathered regarding the completion time, accuracy (when applicable), and replacement for the different interaction schemes.} 
\end{table}

\subsection{Qualitative Evaluation}
We also asked participants general qualitative questions regarding favorite and least favorite methods per interaction scheme (excluding \emph{Maserama Selection}). For \emph{Point Placement}, the method most often ranked as favorite was \emph{Controller-only} (6/17 times), and the strategy most often ranked as least favorite was \emph{Eye-dwell} (8/17 times).  For \emph{Annotation}, the strategy most often ranked as favorite was \emph{Voice recording} (9/17 times), and the one ranked as least favorite was \emph{Handwriting} (8/17 times). In regards to \emph{Maserama}, 12 of the 17 participants found it easy and intuitive to use.

\section{Discussion}\label{sec4}
In our study, the controller and head pointing techniques allow a point to be placed under the surface of an aneurysm while gaze-based methods require ray tracing and thus can only annotate points on the surface. Thus, we defined a “perfect placement” as a point that resides either within or on the surface of the aneurysm. At times, gaze-based interactions can lead to point placement on the adjacent blood vessels, making their accuracy slightly lower than the other types. In terms of efficiency (task completion time) and physical demand, the included strategies are similar. Interestingly, this is contrary to the expectation that eye-tracking could reduce motor efforts and improve efficiency~\cite{bib22}. The number of point replacements before confirmation was 0.8±1.1 (Controller-only), 0.3±0.4 (\emph{Head pointing \& Controller}), 2.1±1.9 (\emph{Gaze \& Controller}), and 5.8±4.5 (\emph{Eye-dwell}). The first two strategies each averaged less than one re-placement per confirmation, and the eye-tracking types were higher. This may be explained by the users’ familiarity with controller-based operations and the instability of gaze-tracking due to microsaccades. With the total NASA TLX score, we revealed that on average, \emph{Controller-only} interaction has the best usability, but the overall performance is not significantly different from \emph{Head pointing \& Controller} and \emph{Gaze \& Controller} methods. Finally, \emph{Eye-dwell} is consistently the worst interaction approach, likely because using the wait time as a trigger to confirm a point adds mental and temporal demand, leading to high frustration. Instead of 3D point placement in complex anatomical models, previous related methodology assessments ~\cite{bib10, bib23, bib24, bib25, bib26} mainly focused on object selection tasks involving 2D menus and simplified shapes and comparing fewer interaction techniques. In general, our results corroborate the overall recommendation from earlier reports that \emph{Controller-only} interaction is preferred over eye-tracking-based methods ~\cite{bib10, bib24}. Furthermore, between \emph{Controller-only} and \emph{Gaze \& Controller}, Luro et al. ~\cite{bib12} showed that while \emph{Gaze \& Controller} is less accurate, its usability and efficiency are on par with the \emph{Controller-only} method. Our results agree with their observations while comparing other additional interaction methods at the same time. When it comes to choosing among \emph{Head pointing \& Controller}, \emph{Gaze \& Controller}, and \emph{Eye-dwell}, the recommendations are not consistent across different studies ~\cite{bib23, bib24, bib25, bib26}. Hou et al. ~\cite{bib10} also observed that the differences between these VR selection techniques are weaker when interacting with 3D objects with multiple depths than simpler 2D menus. In our case, we found a significant difference between \emph{Head pointing \& Controller} and \emph{Eye-dwell} in total NASA TLX. Although many ~\cite{bib22} admit the intuitiveness of eye-tracking-based selection methods, the related technology still requires more investigation to demonstrate its full potential, and mixing with other motor-based interaction schemes (e.g., controller/head motion) ~\cite{bib27} may help better enhance the value of gaze-based interaction methods. 

For the Annotation task, the NASA TLX indicated that \emph{Handwriting} performed worse than the others (p$<$0.05). While \emph{Voice recording} demonstrated lower average workload than \emph{Virtual keyboard}, they only differ in terms of physical demand. To adapt to different annotation needs, we tested annotation with both short and long phrases. The mean completion time for \emph{Voice recording} annotation was 32 seconds, for both long and short phrases. This is comparable to the short phrase versions for \emph{Handwriting} and \emph{Virtual keyboard}, but is about 10 seconds faster than their long phrase versions on average. All annotation strategies were comparable in terms of the mean number of revisions per confirmation, varying between 0.1 and 0.3. Voice recording is also an optimal candidate when it comes to task load, with the lowest average demand.

While the task completion times between long and short phrases have a larger disparity for \emph{Handwriting} and \emph{Virtual keyboard}, such difference is not as striking for \emph{Voice Recording}. This can be explained by the fact that spoken words require much less time to be uttered than to be written or typed. Both \emph{Voice Recording} and \emph{Virtual keyboard} techniques have their own pros and cons. While the first takes far less effort to produce for the annotators, it can take time and be complicated to “read” the notes, particularly for long messages with non-verbal symbols. The latter takes time to produce, but is more beneficial to the note reviewers. It is possible that with more intelligent keyboard inputting techniques (i.e., automatic word completion), the \emph{Virtual keyboard} method may be more advantageous. In current literature~\cite{bib28}, there is still a lack of comprehensive comparison of VR note-taking techniques across different input modalities, and existing assessments primarily focus on variants of keyboard-based techniques (physical or virtual). Among virtual keyboard techniques that use different ways to select and confirm the key, key selection with Controllers (the same technique tested in our study) outperforms the rest ~\cite{bib29, bib30}. To further improve the performance, studies~\cite{bib28} have suggested using a 3D virtual keyboard instead of a planar type and alternative key arrangement, such as circular designs. Although not tested in our study, incorporating physical keyboards in virtual views has also been attempted~\cite{bib28}, but challenges still exist, such as required attention shift and small size of the physical keys. Finally, voice-based texting~\cite{bib31} showed higher efficiency than keyboard entries on touchscreen phones, and in mixed reality, it also demonstrated sufficient performance against moderate environmental noise~\cite{bib32}. However, different from our implementation, existing techniques~\cite{bib28} often incorporate speech dictation algorithms, which can introduce errors due to the state-of-the-art for the relevant methods.

Our newly proposed \emph{Maserama} selection technique received positive feedback from the user study. The design of the technique was inspired by Magnoramas (Magnified Dioramas)~\cite{bib16} which employs a magnified copy of a virtual object for more precise annotation in the context of multi-party VR collaboration. Different from Magnorama that was demonstrated for a single full anatomical model (i.e., a skull), Maserama’s novelty comes from its sub-volume selection function and dynamic cropping features. Participants were able to reselect spherical volumetric cropping of the target at will, and later had the ability to select individual structures within their sliced cluster. This is more beneficial for complex anatomical models. With a fast completion time of 21.0±9.3 seconds, it is a highly efficient VR interaction method. In terms of robustness, two different types of selection revisions were gathered, including the number of re-selection of sub-volumes and that of target structure selection. While the first was measured at 0.4 on average per person per task, the second metric was averaged at 0.04 per person per task. These low revision counts further solidify its high robustness and usability. In terms of task load, it scored the second lowest among all interaction strategies, with 10.2±9.9. Based on the feedback from the participants, further adjustments can be made to enhance its user-friendliness, including allowing user controls of  spherical volume size and virtual rod length. We will incorporate these features in the future system and validate it through additional user studies. 

We have identified the most effective and user-friendly VR interactive schemes for anatomical data in the tasks of 3D point placement and annotation, namely \emph{Controller-only} and \emph{Voice recording}. These can also be combined with our newly proposed \emph{Maserama} method to form a new workflow, where precise point location can be tagged and well annotated for the desirable structures after they have been magnified and isolated through \emph{Maserama} selection. With each component tested individually already, we will further examine the robustness and usability of the complete workflow, especially for multi-party medical VR collaboration that requires more complex setups and considerations in the future.

One limitation of this study lies in the lack of diversity of anatomical variability and limited participants, including clinicians. While additional anatomical models can further improve the reliability of the evaluations, they will also prolong the experiments causing fatigue to the participants, adversely affecting the assessment. With well defined objectives for each experiment, non-clinician users were able to complete the tasks efficiently to yield meaningful insights and comparisons. We believe the same will hold true for clinicians. However, due to the global pandemic, recruiting clinicians was challenging. Future works will expand on the experiments with more complex tasks, data, and evaluations. We will expand our evaluations for more application-specific context with a larger cohort of participants, especially clinical specialists  in the future.

\section{Conclusion}\label{sec5}
We have evaluated a number of user-interaction strategies for navigation and annotation of neurosurgical data in VR. More specifically, we compared four interaction techniques for point placement and three for note-taking. While our studies invalidated our hypothesis for point placement by instead suggesting \emph{Controller-only} interaction as the preferred technique among all, including those that integrate eye-tracking technology, the initial hypothesis regarding the best technique for note-taking is supported by the experimental outcomes. This lack of gaze effectiveness is unexpected, and is thus valuable information. It demonstrates that eye-tracking is usable in the neurosurgical VR context, but calls for further investigation on how gaze information can be better integrated for medical VR applications. Furthermore, we have also proposed and validated a novel data interaction technique, called \emph{Maserama}, for sub-volume exploration and selection, and its efficiency and usability are confirmed by quantitative and qualitative assessments. In summary, our studies provide the needed insights in designing HCI methods for medical VR applications, where complex 3D anatomical data pose the main challenges. They contribute meaningful evaluation of a range of techniques, by validating their usefulness, highlighting their differences, and presenting most effective strategies.

\backmatter

%\bmhead{Supplementary information}

\bmhead{Statements and Declarations}
The authors declare no financial or non-financial interests that are directly or indirectly related to the work submitted for publication

\bmhead{Data availability}
The datasets generated during and/or analysed during the current study are available from the corresponding author on reasonable request.

\bibliography{sn-bibliography}% common bib file
%\input sn-article.bbl
%\input sn-bibliography.bib

%\input sn-sample-bib.tex%

\end{document}